\documentclass[12pt]{iopart}
\usepackage{iopams}
\usepackage{graphicx}
\begin{document}

\title[]
{Dynamic localization of lattice electrons under time dependent
electric and magnetic fields}

\author{C. Micu$^\dag$ E. Papp$^\ddag$ and L. Aur$^\ddag$}

\address{\dag\ Physics Department, North University of Baia Mare, RO-430122,
Baia Mare, Romania}

\address{\ddag\ Department of Theoretical
Physics, West University of Timisoara, RO-300223, Romania}

\begin{abstract}
Applying the method of characteristics leads to wavefunctions and
dynamic localization conditions for electrons on the one dimensional
lattice under perpendicular time dependent electric and magnetic
fields. Such conditions proceed again in terms of sums of products
of Bessel functions of the first kind. However, this time one deals
with both the number of magnetic flux quanta times $\pi $ and the
quotients between the Bloch frequency and the ones characterizing
competing fields. Tuning the phases of time dependent modulations
leads to interesting frequency mixing effects providing an
appreciable simplification of dynamic localization conditions one
looks for. The understanding is that proceeding in this manner, the
time dependent superposition mentioned above gets reduced
effectively to the influence of individual ac-fields exhibiting
mixed frequency quotients. Besides pure field limits and
superpositions between uniform electric and time dependent magnetic
fields, parity and periodicity effects have also been discussed.
\end{abstract}

%\maketitle

\vspace{2pc} \noindent\textit{Keywords}: Mesoscopic and nanoscale
systems; Superlattices; Localization effects. Submitted to: J.Phys.
: Condens. Matter

\section{Introduction}

The dynamic localization (DL) of electrons moving on the one
dimensional (1D) lattice under the influence of a longitudinal time
dependent (TD) electric field like $E(t)=E_{0}f(t)$ has received
much interest since its discovery some 20 years ago [1-8]. This
effect concerns the periodic return of the electron to the initially
occupied site [1]. Accordingly, the mean square displacement (MSD)
should remain bounded in time. In most cases which
are of interest in practice the modulation function is periodic with period $%
T$, but superpositions of several ac-fields can also be considered.
The DL referred to above should then occur under selected conditions
concerning the quotients between the field frequencies and the Bloch
frequency $\omega _{B}=E_{0}ea/\hbar $, where $a$ stands for the
lattice spacing. Besides applications in several areas like high
field and nonlinear effects [2,3], trapping in two level atoms [4],
persistent THz emission [5,6], the generation of higher harmonics
[7] or the absolute negative conductance, the DL has been finally
observed in the linear optical absorption coefficient of quantum dot
superlattices [9]. It has also been found that the collapse of
quasienergy bands is able to reflect the occurrence of DL [10-13].
Recent developments such as the DL of wave packets in barriers with
TD parameters [14], the appearance and disappearance of resonant
peaks in $I-V$ characteristics [15], or the influence of higher
order neighbors [16], are worthy of being mentioned, too.

We have to realize that the influence of the uniform magnetic field
on lattice electrons, such as exhibited by the celebrated
Harper-equation [17-19], looks quite interesting in many respects.
The same concerns several superpositions between electric and
magnetic fields which have been studied during course [20-22].
However, a systematic study of wavefunctions and of DL effects
produced by superpositions for which both electric and magnetic
fields are TD seems of having been, to the best of our knowledge,
overlooked. We shall then use this opportunity to discuss DL effects
provided by the longitudinal TD electric field $\overrightarrow{E}%
(t)=(E(t),0,0)$ working in conjunction with a transversal magnetic
field like $\overrightarrow{B}(t)=(0,0,B(t))$, where
$B(t)=B_{0}g(t)$. To this aim we can start either by incorporating
the electric field into the time dependent Harper-Hamiltonian [18],
or from the decoupled limit of two parallel chains in electric and
magnetic fields [21]. We have to realize that such systems can be
converted one into another with the help of gauge transformations
[23]. The former alternative is appropriate for Hall conductance
studies. We shall then choose the latter alternative since it
complies in a more suitable manner with the DL problem. This opens
the way to the derivation of general DL conditions, but concrete
realizations such as superpositions between uniform electric and
time dependent magnetic fields, will also be discussed. In this
context, we found that tuning the phases of TD modulations opens the
way to interesting frequency mixing effects, which leads in turn to
quickly tractable DL conditions proceeding again in terms of the
zero's of the zero-order Bessel function of the first kind. The
understanding is that by virtue of the phase tuning, the TD
superposition mentioned above is able to be reduced effectively to
the influence of individual ac-fields characterized by mixed
frequency quotients. Last but not at least we shall deal with parity
and periodicity effects.

\section{The derivation of the wavefunction}

The Hamiltonian describing the electron on the 1D lattice under TD
electric and magnetic fields specified above is given by [1,18,21]

\begin{eqnarray*}
\mathcal{H}=\varepsilon _{0}\sum\limits_{m}\mid m><m\mid
-eaE_{0}f(t)\sum\limits_{m}m\mid m><m\mid +
\end{eqnarray*}
\begin{equation}
\{
V%
%TCIMACRO{\dsum \limits_{m}}%
%BeginExpansion
{\displaystyle\sum\limits_{m}}
%EndExpansion
\exp(-i\gamma/2)|m+1><m|+V%
%TCIMACRO{\dsum _{m}}%
%BeginExpansion
{\displaystyle\sum_{m}}
%EndExpansion
\exp(i\gamma/2)|m><m+1| \}
\end{equation}
where $m$ is an integer ranging from $-\infty $ to $\infty $. One
has

\begin{equation}
\gamma =\gamma (t)=2\pi \frac{\Phi }{\Phi _{0}}=2\pi \beta _{0}g(t)
\end{equation}%
where the transversal magnetic flux and the magnetic flux quantum
are given
by $\Phi =B_{0}a^{2}g(t)$ and $\Phi _{0}=hc/e$, respectively. So, there is $%
\beta _{0}=eB_{0}a^{2}/hc$, which stands for the magnetic
commensurability parameter. The constant on-site energy is denoted
by $\varepsilon _{0}=\hbar \omega _{0}$, whereas $V=\hbar U$ is
responsible for the nearest neighbor hopping parameter. One deals,
of course, with an orthonormalized Wannier-basis for which $<m\mid
m^{\prime }>=\delta _{m,m^{^{\prime }}}$. We then have to account,
as usual, for the TD single particle amplitude via

\begin{equation}
\mid \Psi (t)>=\sum\limits_{m}C_{m}(t)\exp (-i\omega _{0}t)\mid m>
\end{equation}%
which leads in turn to the TD second-order discrete Schr\"{o}dinger
equation

\begin{equation}
i\frac{\partial }{\partial t}C_{m}(t)=U\exp (-i\gamma
/2)C_{m-1}(t)+U\exp (i\gamma /2)C_{m+1}(t)-m\omega
_{B}f(t)C_{m}(t)\quad .
\end{equation}%
The next step is to apply the discrete Fourier-transform

\begin{equation}
C_{m}(t)=\frac{1}{2\pi }\int\limits_{0}^{2\pi }dk\exp (imk)\widetilde{C}%
_{k}(t)
\end{equation}%
where $k$ denotes the dimensionless wave number. Then (4) becomes

\begin{equation}
\left[ \frac{\partial }{\partial t}+2iU\cos \left( k+\frac{\gamma }{2}%
\right) +\omega _{B}\widetilde{f}(t)\frac{\partial }{\partial
k}\right] \widetilde{C}_{k}(t)=0
\end{equation}%
where

\begin{equation}
\widetilde{f}(t)=f(t)-\frac{1}{2\omega _{B}}\frac{d\gamma (t)}{dt}=f(t)-%
\frac{\pi \beta _{0}}{\omega _{B}}\frac{dg(t)}{dt}\quad .
\end{equation}

Now we have to remember that (4) and (6) have been discussed before when $%
\gamma =0$ by resorting to the method of characteristics [1]. What
then remains is to generalize these latter results towards
incorporating $\gamma (t)$ such as given by (2). This results in the
solution

\begin{equation}
\widetilde{C}_{k}(t)=\exp \left[ -2iU\int\limits_{0}^{t}dt^{\prime
}\cos \Omega _{k}(t,t^{\prime })\right]
\end{equation}%
as it can be easily verified by direct computation. This time one
has

\begin{equation}
\Omega _{k}(t,t^{\prime })=\cos \left[ k+\frac{\gamma }{2}-\omega
_{B}\left( \widetilde{\eta }(t)-\widetilde{\eta }(t^{\prime
})\right) \right]
\end{equation}%
where

\begin{equation}
\widetilde{\eta }(t)=\int\limits_{0}^{t}dt^{\prime
}\widetilde{f}(t^{\prime })=\eta (t)-\frac{1}{2\omega _{B}}\left(
\gamma (t)-\gamma (0)\right)
\end{equation}%
and

\begin{equation}
\eta (t)=\int\limits_{0}^{t}dt^{\prime }f(t^{\prime })
\end{equation}%
which is well known in the description of electric field problems.
Accordingly, one gets faced with the normalized wavefunction

\begin{equation}
C_{m}(t)=\exp (-im\widetilde{\psi })J_{m}(2U\mid
\widetilde{Z}(t)\mid )
\end{equation}%
by virtue of the well known properties of Bessel functions [24],
where

\begin{equation}
\widetilde{\psi }=\widetilde{\psi }(t)=\frac{\pi +\gamma
(t)}{2}-\arg \widetilde{Z}(t)
\end{equation}%
and

\begin{equation}
\widetilde{Z}(t)=\int\limits_{0}^{t}dt^{\prime }\exp (-i\omega _{B}%
\widetilde{\eta }(t^{\prime }))\quad
\end{equation}%
which deserve further attention.

.

\section{Dynamic localization effects}

\bigskip Using (12) produces the MSD

\begin{equation}
<m^{2}>=\sum\limits_{m}m^{2}\mid C_{m}(t)\mid ^{2}=2U^{2}\mid \widetilde{Z}%
(t)\mid ^{2}
\end{equation}%
which generalizes apparently (2.7) in [1] in terms of the
substitution $\eta (t)\rightarrow $ $\widetilde{\eta }(t)$. Choosing
as an example usual modulations like

\begin{equation}
f(t)=\cos (\omega _{1}t)
\end{equation}%
and

\begin{equation}
g(t)=\sin (\omega _{2}t)
\end{equation}%
yields the characteristic function

\begin{equation}
\widetilde{Z}(t)=\int\limits_{0}^{t}dt^{\prime }\exp \left(
-i\frac{\omega _{B}}{\omega _{1}}\sin (\omega _{1}t^{\prime })+i\pi
\beta _{0}\sin (\omega _{2}t^{\prime })\right)
\end{equation}%
which can be rewritten equivalently as

\begin{equation}
\widetilde{Z}(t)=\sum\limits_{m}\sum\limits_{n}\exp \left( i\Omega
_{m,n}t\right) \frac{\sin \left( \Omega _{m,n}t\right) }{\Omega _{m,n}}%
J_{n}\left( \frac{\omega _{B}}{\omega _{1}}\right) J_{m}\left( \pi
\beta _{0}\right)
\end{equation}%
by virtue of expansions characterizing generating functions of
Bessel functions [24], where

\begin{equation}
\Omega _{m,n}=\frac{1}{2}\left( m\omega _{2}-n\omega _{1}\right)
\quad
\end{equation}%
and $q_{j}=\omega _{B}/\omega _{j}$ $(j=1,2)$. The point is to decompose $%
\widetilde{Z}(t)$ as

\begin{equation}
\widetilde{Z}(t)=Q_{1}t+Q_{2}(t)
\end{equation}%
in which $Q_{2}(t)$ oscillates with time. Having discriminated the
linear term in $t$ then produces the DL condition [1,13]

\begin{equation}
Q_{1}=0
\end{equation}%
in which case the MSD remains bounded in time. On the other hand
(19) shows that the discrimination of the linear term one looks for
proceeds in terms of selected $m$- and $n$-values for which $\Omega
_{m,n}\rightarrow 0$. To this aim let us assume that the frequencies
$\omega _{1}$ and $\omega _{2}$ are commensurate. This amounts to
deal with quotients like

\begin{equation}
\frac{n}{m}=\frac{\omega _{2}}{\omega
_{1}}=\frac{q_{1}}{q_{2}}=\frac{P}{Q}
\end{equation}%
in which $P$ and $Q$ are mutually prime integers. We then have to
realize that the DL condition characterizing specifically the
present superposition of TD electric and magnetic fields is given in
terms of sums of products of Bessel functions of the first kind like

\begin{equation}
Q_{1}=Q_{1}\left( q_{1},\pi \beta _{0}\right) =J_{0}\left(
q_{1}\right) J_{0}\left( \pi \beta _{0}\right)
+\sum\limits_{l=1}^{\infty }p_{l}J_{Pl}\left( q_{1}\right)
J_{Ql}\left( \pi \beta _{0}\right) =0
\end{equation}%
in which

\begin{equation}
p_{l}=1+(-1)^{l(P+Q)}
\end{equation}%
and $P=Q\omega _{2}/\omega _{1}$. Using, however,
$\widetilde{g}(t)=\cos (\omega _{2}t)$ instead of (17) produces the
DL condition

\begin{equation}
\widetilde{Q}_{1}\left( q_{1},\pi \beta _{0}\right) =\left[
J_{0}\left(
q_{1}\right) J_{0}\left( \pi \beta _{0}\right) +\sum\limits_{l=1}^{\infty }%
\widetilde{p}_{l}J_{Pl}\left( q_{1}\right) J_{Ql}\left( \pi \beta
_{0}\right) \right] =0
\end{equation}%
which proceeds up to a phase factor like $\exp (-i\pi \beta _{0})$,
where

\begin{equation}
\widetilde{p}_{l}=\exp \left( i\frac{\pi }{2}Ql\right) +\exp \left( -i\frac{%
\pi }{2}Ql\right) (-1)^{l(P+Q)}\quad .
\end{equation}
Such results indicate that contributions provided by electric and
magnetic
fields can be placed on the same footing. Of course, starting from $%
f(t)=g(t)=\cos (\omega _{1}t)$, one finds that the DL condition is
still given by (26), but this time via $P=Q=1$.

One remarks that both (24) and (26) are sensitive to the parity of
$P+Q$. So (24) becomes

\begin{equation}
Q_{1}=Q_{1}^{(-)}\left( q_{1},\pi \beta _{0};\xi _{l}\right)
=J_{0}\left( q_{1}\right) J_{0}\left( \pi \beta _{0}\right)
+2\sum\limits_{l=1}^{\infty }\xi _{l}J_{2Pl}\left( q_{1}\right)
J_{2Ql}\left( \pi \beta _{0}\right) =0
\end{equation}%
when $P+Q$ is an odd integer, but

\begin{equation}
Q_{1}=Q_{1}^{(+)}\left( q_{1},\pi \beta _{0};\xi _{l}\right)
=J_{0}\left( q_{1}\right) J_{0}\left( \pi \beta _{0}\right)
+2\sum\limits_{l=1}^{\infty }\xi _{l}J_{Pl}\left( q_{1}\right)
J_{Ql}\left( \pi \beta _{0}\right) =0
\end{equation}%
when $P+Q$ is even. So far we have to insert $\xi _{l}=1$, but
further generalizations are in order. Note that the presence of $\xi
_{l}$ in (28) and (29) has to be understood just as an insertion
prescription working under the sum.

Proceeding in a similar manner one finds that (26) leads to
\begin{equation}
\widetilde{Q}_{1}^{(+)}\left( q_{1},\pi \beta _{0}\right)
=Q_{1}^{(+)}\left( q_{1},\pi \beta _{0};\cos (\pi Ql/2)\right) =0
\end{equation}%
and

\begin{equation}
\widetilde{Q}_{1}^{(-)}\left( q_{1},\pi \beta _{0}\right)
=Q_{1}^{(-)}\left( q_{1},\pi \beta _{0};\cos (\pi Ql)\right)
+i\Gamma _{1}=0
\end{equation}%
when $P+Q$ is even and odd, respectively, where

\begin{equation}
\Gamma _{1}=\Gamma _{1}(q_{1},\pi \beta
_{0})=2\sum\limits_{l=1}^{\infty }\sin \left( \frac{\pi
}{2}(2l+1)Q\right) J_{(2l+1)P}\left( q_{1}\right) J_{(2l+1)Q}\left(
\pi \beta _{0}\right) \quad .
\end{equation}%
In the latter case the DL occurs when both real and imaginary parts of $%
\widetilde{Q}_{1}^{(-)}\left( q_{1},\pi \beta _{0}\right) $ are
zero, so that we have to account for twice harder computations. Such
conditions are rather complex, so that when dealing with
applications we have to resort to numerical studies. We can then say
that looking for reasonable but tractable versions represents a
quite desirable task.

\section{Phase tuning and frequency mixing effects}

A further point of interest is to account for the influence of
phases on the DL. This opens the way to the derivation of reasonable
DL conditions one looks for. For this purpose let us start from some
appropriate modulations like

\begin{equation}
f(t)=\cos (\omega _{1}t+\theta _{1})
\end{equation}%
and

\begin{equation}
g(t)=\cos (\omega _{1}t+\theta _{2})
\end{equation}%
for which $\omega _{1}=$ $\omega _{2}$ and $\theta _{j}\in \lbrack 0,2\pi ]$%
. Repeating the same steps as before yields the DL condition

\begin{equation}
\widetilde{Q}_{1}\left( R_{1},R_{2}\right) =\sum\limits_{m=-\infty
}^{\infty }iJ_{m}\left( R_{1}\right) J_{m}\left( R_{2}\right) =0
\end{equation}%
where

\begin{equation}
R_{1}=q_{1}\cos (\theta _{1})+\pi \beta _{0}\sin (\theta _{2})
\end{equation}%
and

\begin{equation}
R_{2}=\pi \beta _{0}\cos (\theta _{2})-q_{1}\sin (\theta _{1})
\end{equation}%
are responsible for mixing effects concerning electric and magnetic
quotients mentioned above. Discriminating the $n=0$-term in (35)
leads to the decomposition

\begin{equation}
\widetilde{Q}_{1}\left( R_{1},R_{2}\right)
=J_{0}(R_{1})J_{0}(R_{2})+\Delta \widetilde{Q}_{1}
\end{equation}%
where

\begin{equation}
\Delta \widetilde{Q}_{1}=2\sum\limits_{n=1}^{\infty }\cos (\pi
n/2)J_{n}(R_{1})J_{n}\left( R_{2}\right) \quad .
\end{equation}

Fixing, for convenience, the $\theta _{1}$-phase, one sees that the
$\theta _{2}$-phase can be tuned until $R_{2}=0$, in which case

\begin{equation}
\cos (\theta _{2})=\frac{q_{1}}{\pi \beta _{0}}\sin (\theta
_{1})\quad .
\end{equation}%
Accordingly, (38) becomes, interestingly enough,

\begin{equation}
\widetilde{Q}_{1}\left( R_{1},0\right)
=J_{0}(\widetilde{R}_{1})=0\quad .
\end{equation}%
This shows that the celebrated DL condition derived before gets
reproduced [1], now in terms of the mixed frequency quotient

\begin{equation}
\widetilde{R}_{1}=q_{1}\cos (\theta _{1})+\pi \beta _{0}\left[ 1-\frac{%
q_{1}^{2}}{\pi ^{2}\beta _{0}^{2}}\sin ^{2}(\theta _{1})\right]
^{1/2}\quad
\end{equation}%
relying effectively on the influence of a single ac-field. So one
gets faced
with the simplified DL condition $\widetilde{R}_{1}=z_{n}$, where $z=z_{n}$ (%
$n=1,2,3,...$) stands for the root of $J_{0}(z)$. This condition can
be rewritten equivalently as

\begin{equation}
z_{n}^{2}+2z_{n}q_{1}\sin (\theta _{1})+q_{1}^{2}=\pi ^{2}\beta
_{0}^{2}
\end{equation}%
which is able to serve as an eigenvalue equation for $\pi \beta _{0}$ or $%
q_{1}$, respectively. However, the $\theta _{2}$-phase can also be
tuned so that $\pi \beta _{0}=0$, in which case the DL condition
becomes

\begin{equation}
\widetilde{Q}_{1}\left( 0,R_{2}\right)
=J_{0}(\widetilde{R}_{2})=0\quad
\end{equation}%
instead of (41), where now the frequency quotient is given
effectively by

\begin{equation}
\widetilde{R}_{2}=-q_{1}\sin (\theta _{1})+\pi \beta _{0}\left[ 1-\frac{%
q_{1}^{2}}{\pi ^{2}\beta _{0}^{2}}\cos ^{2}(\theta _{1})\right]
^{1/2}\quad .
\end{equation}%
One sees immediately that $\widetilde{R}_{2}=z_{n^{\prime }}$ by
virtue of (44), so that (43) is replaced by

\begin{equation}
z_{n^{\prime }}^{2}-2z_{n^{\prime }}q_{1}\cos (\theta
_{1})+q_{1}^{2}=\pi ^{2}\beta _{0}^{2}
\end{equation}%
where, in general, $n^{\prime }$ differs from $n$. We then have to
account for two realizations of the DL condition, such as indicated
by (43) and
(46). One realizes that dealing with $\widetilde{R}_{1}$ and $\widetilde{R}%
_{2}$ amounts to introduce effectively the influence of individual
fields selected in accord with (42) and (45). However, we can choose
$\theta _{1}$ such that

\begin{equation}
\sin (\theta _{1})+\cos (\theta _{1})=0
\end{equation}%
which also means that $\theta _{1}\in \lbrack \pi /2,\pi ]$ or
$\theta _{1}\in \lbrack 3\pi /2,2\pi ]$. The corresponding DL
condition is then given by

\begin{equation}
\widetilde{R}_{1}=\widetilde{R}_{2}=z_{n}
\end{equation}%
where by now $n^{\prime }=n$, which represents an appreciable
simplification. One realizes that (48) produces quantized flux
values like

\begin{equation}
\beta _{0}=\beta _{0}(n)=\frac{1}{\pi }\left[
z_{n}^{2}-2z_{n}q_{1}\cos (\theta _{1})+q_{1}^{2}\right] ^{1/2}
\end{equation}%
when starting from fixed $q_{1}$-quotients. Such flux values exhibit
unexpected limits like

\begin{equation}
\pi \beta _{0}(n)\rightarrow z_{n}\pm q_{1}
\end{equation}%
if $\theta _{1}\rightarrow \pi /2$ and $\theta _{1}\rightarrow 3\pi
/2$, respectively.

\section{Other concrete realizations}

The superposition between a uniform electric field for which
$f(t)=1$ and a TD magnetic one deserves a little bit more attention.
This time (10) becomes

\begin{equation}
\widetilde{\eta }(t)=t-\frac{\pi \beta _{0}}{\omega _{B}}\left(
g(t)-g(0)\right)
\end{equation}%
so that

\begin{equation}
\widetilde{Z}(t)=\sum\limits_{m}\exp \left( i\Omega _{m}t\right)
\frac{\sin \left( \Omega _{m}t\right) }{\Omega _{m}}J_{n}\left(
q_{1}\right) J_{m}\left( \pi \beta _{0}\right)
\end{equation}%
where now

\begin{equation}
\Omega _{m}=\frac{1}{2}\left( m\omega _{2}-\omega _{B}\right) \quad
.
\end{equation}%
One sees that $\widetilde{Z}(t)$ contains only oscillatory
contributions in so far as $\omega _{B}<\omega _{2}$. This means in
turn that $Q_{1}=0$, so that requirements needed for the onset of DL
are fulfilled from the very beginning. Furthermore let us assume
that

\begin{equation}
\omega _{B}=\widetilde{n}\omega _{2}
\end{equation}%
if $\omega _{B}\eqslantgtr \omega _{2}$, where $\widetilde{n}$
denotes a positive integer. Then the linear term gets discriminated
again via $\Omega _{m}\rightarrow 0$, which leads to a rather
special DL condition like

\begin{equation}
Q_{1}=Q_{1}^{\ast }\left( q_{2},\pi \beta _{0}\right) =\theta \left(
q_{2}-1\right) J_{\widetilde{n}}(\pi \beta _{0})=0
\end{equation}%
where $\theta (x)$ stands for Heaviside's function. One realizes
that this time the DL proceeds selectively in terms of zero's
characterizing higher order Bessel functions for which
$\widetilde{n}=1,2,3,...$, in accord with (32). Complementarily, the
regime is ballistic when $J_{\widetilde{n}}(\pi \beta _{0})\neq 0$,
but the DL gets restored again when the quotient $\omega _{B}/\omega
_{2}$ is either pure rational or irrational.

Next let us account for pure electric and magnetic fields via $\beta
_{0}\rightarrow 0$ and $\omega _{B}\rightarrow 0$, respectively. In
the first case one recovers the well-known DL condition [1]

\begin{equation}
Q_{1}\left( q_{1}\right) =J_{0}\left( q_{1}\right) =0\quad .
\end{equation}%
In the second case one finds

\begin{equation}
Q_{1}\left( 0,\pi \beta _{0}\right) =J_{0}(\pi \beta _{0})=0
\end{equation}%
which means that the DL occurs whenever the number of flux quanta
times $\pi $ is a root of the zero order Bessel function, too. This
proceeds in terms of dimensionless flux values centered around
$\beta _{0}=\beta _{n}\cong z_{n}/\pi $, where $J_{0}(z_{n})=0$. The
approximation

\begin{equation}
\beta _{n+1}-\beta _{n}\cong 1
\end{equation}%
should also be mentioned, which indicates that the DL occurs
periodically with unit dimensionless flux period. A such
periodicity, which is reminiscent to Aharonov-Bohm oscillations, has
also been remarked in superpositions between uniform magnetic and
dc-ac electric fields [22].

\section{Conclusions}

The influence of TD electric and magnetic fields on the DL of
electrons on the 1D lattice has been discussed systematically in
terms of MSD's remaining bounded in time. The main result is given
by the rather general formula (24). A such formula is able to
exhibit several concrete realizations and serves as a starting point
for further developments. Indeed, tuning the phases results in
interesting frequency mixing effects which provide controllable DL
conditions working again in terms of the zero-order Bessel function,
as shown by (41) and (44). For this purpose one resorts to mixed
frequency quotients such as given by (42) and (45). Such quotients
proceed effectively in terms of individual ac-fields. Related flux
quantization rules have also been derived in accord with (49), now
by starting from fixed values of the electric frequency quotient
$q_{1}=\omega _{B}/\omega _{1}$. Of course, (41) and (44) can also
be discussed by starting from fixed $\pi \beta _{0}$-values, which
leads in turn to the quantization of the electric field amplitude.
We emphasize that such results can be readily generalized towards
modulations for which $\omega _{1}\neq \omega _{2}$. We have also to
remark that (49) can be rewritten in terms of pertinent vectors as

\begin{equation}
\pi \overrightarrow{\beta }_{0}(n)+\overrightarrow{q}_{1}=\overrightarrow{z}%
_{n}
\end{equation}%
with the understanding that $\theta _{1}$ stands for the angle between $%
\overrightarrow{z}_{n}$ and $\overrightarrow{q}_{1}.$

Pure fields limits can be readily performed, as shown by (56) and
(57). One realizes that the first root in (57), namely the selected
dimensionless magnetic flux

\begin{equation}
\beta _{0}=\beta _{1}\cong \frac{2.405}{\pi }\cong 0.76
\end{equation}%
deserves experimental verification in a close analogy with the
confirmation of the DL concerning the "electric" quotient $\omega
_{B}/\omega _{1}=z_{1}\cong 2.405$ [9]. The superposition between a
uniform electric field and a TD magnetic one has also been
discussed. This is a rather special example for which the fields are
accounted for in an asymmetric manner. Now the DL occurs in terms of
(55), but when the Bloch frequency is quantized in accord with (54)
only. Correspondingly, the electric field should be quantized itself
by virtue of the rule

\begin{equation}
E_{0}=E_{0}^{(n)}=\widetilde{n}\frac{\hbar \omega _{2}}{ea}\quad
\end{equation}%
which can be viewed as being similar to (49). We can then say that
results presented in this paper provide a deeper understanding of DL
effects, with a special emphasis on the role of phase tuning
effects. Applications in the design of nanoelectronic devices can
also be invoked.

\section*{Acknowledgement}

We are indebted to CNCSIS/Bucharest for financial support

\end{document}